\documentclass[graybox]{svmult}

\usepackage{helvet}         
\usepackage{courier}        
\usepackage{type1cm}        
\usepackage{makeidx}         
\usepackage{graphicx}        
\usepackage{multicol}        
\usepackage[bottom]{footmisc}
\usepackage{amsmath ,amssymb}
\usepackage{bbold}
\usepackage{stmaryrd}
\usepackage{hyperref}

\begin{document}

\title*{Dual dilaton with $\mathcal{R}$ and $\mathcal{Q}$ fluxes}
\author{Eugenia Boffo}
\institute{E Boffo \at Charles University Prague, Karl\'{i}n, Sokolovsk\'{a} 83,\\ \email{boffo@karlin.mff.cuni.cz}}
%
%
\maketitle

\abstract{In previous works we showed that a Courant algebroid in a particular frame and the differential geometry of the sum bundle $TM \oplus T^*M$ provide a very natural geometric setting for a sector of the low energy effective limit of type II superstring theories (Supergravity theory). Given our geometric and algebraic considerations, we reproduced the NS-NS sector of the closed bosonic effective type II sting action, and an action for the inverse metric $G^{-1}$ and the bivector $\Pi$, related to the tensors for closed strings as $(g+B)^{-1} = (G^{-1} +\Pi)$. The action depended on the stringy T-dual fluxes $\mathcal{R}$ and $\mathcal{Q}$, but the dual dilaton was missing. This short paper fills the gap.}

\bigskip

Contribution to the proceedings of the XIV International Workshop "Lie Theory and its Applications in Physics", 20-25 June 2021, Sofia, Bulgaria.

\section{Introduction}
\label{sec:1}

The geometric setting for string effective actions and ultimately Supergravity is notably the generalized tangent bundle $TM \oplus T^* M$, as recognized in \cite{Coimbra:2011nw}. Geometric actions with fluxes were later constructed in the context of Generalized Geometry and Double Field Theory \cite{Andriot:2013xca}, \cite{Andriot:2012an} and \cite{Andriot:2012wx}. The results acquired a further meaning thanks to the Lie algebroid arguments of \cite{Blumenhagen:2012nk} and \cite{Blumenhagen:2013aia}. In a recent work \cite{Boffo:2020} we suggested a curvature scalar for the target space metric seen by the open strings. The T-dual fluxes $\mathcal{R}$ and $\mathcal{Q}$ were naturally encoded in the expression for the curvature scalar. To derive our result in \cite{Boffo:2020} we relied not only on a Courant algebroid and on the differential geometry of the sum bundle $TM \oplus T^*M$, but also on the canonical correspondence with a dg-symplectic manifold of degree $2$. Relevance was given to the graded description as it allows for a neat extrapolation of the underlying $2$-cocycle structure of the Courant algebroid (the 1-gerbe $B\in \Omega^2(M)$). To keep the current paper short we will not present this viewpoint here. Moreover, in the graded Poisson structure it is believed that the dilaton should account for an ambiguity in the quantization of the sheaf of graded functions \cite{Gualtieri2004}.

The construction of the curvature invariant goes as follows.
Requiring that the space of sections of the direct sum bundle $TM \oplus T^*M$ could have two bilinear products, namely a Courant algebroid bracket and a Lie-type of bracket, is enough to leave an affine metric connection, with totally skew-symmetric torsion, completely defined. A connection on sections of some subbundles, in particular $1$-forms in $\Gamma(T^*M)$, can be defined too: One must just ask that a splitting $r : T^*M \mapsto TM \oplus T^*M$ of the short exact sequence $ 0 \mapsto TM \hookrightarrow TM \oplus T^*M \twoheadrightarrow T^*M \mapsto 0$ can be a morphism between vector bundles with connections, and apply it to the fully fledged connection on generalized tangent vectors. The curvature scalar of this "smaller" connection is retained in the usual fashion, i.e.\,by contracting the free indices of the commutator of covariant derivatives. In a similar way, with a split $s$ of the exact Courant algebroid sequence, $s: TM \mapsto TM \oplus T^*M$ and with the associated connection on vector fields, the related curvature scalar can reproduce the Lagrangian for the NS-NS closed bosonic sector of the effective string action, if the appropriate choices are made.

In the body of the article we will apply this method starting with a particular choice of frame that depends on all the relevant fields, especially the dual dilaton, so far missing.

\section{Dual dilaton}
\label{sec:2}
A dual dilaton can be introduced by requiring that the volume form rescaled with the dilaton remains invariant \cite{Andriot:2012wx}, according to the formula: 
\begin{equation}e^{-2\phi} \sqrt{\det g} \, \mathrm{d}x = e^{-2\tilde{\phi}} \sqrt{\det G^{-1}} \, \mathrm{d}x. \label{dil} \end{equation}
An old lemma by J.\,Moser \cite{Moser1965} states that this is the case if one can find a diffeomorphism of a $n$-dimensional compact oriented manifold with itself, for which the $n$-cycles are preserved. We will heavily rely on the relation \eqref{dil} in our derivation.

\subsection{The Courant algebroid and its connection}
Let us implement the observation about the dual dilaton in a vielbein $\mathcal{E}$ for $TM \oplus T^* M$, as displayed below:
\begin{equation}
	\mathcal{E} = e^{\kappa \phi} \begin{pmatrix}
		 \mathbb{1} & \gamma (G^{-1}-\Pi) \\
		 -(g+B) & \gamma \mathbb{1}
	\end{pmatrix}, \quad \kappa \in \mathbb{R},
\end{equation}
where $g(x) \in S^2(M)$ and non-degenerate, $B(x)\in \Omega^2(M)$ and $G^{-1}(x)-\Pi(x) \equiv (g(x)-B(x))^{-1}$ (open-closed string metrics relations, where $G^{-1}$ is the symmetric part and $\Pi$ a bivector). Besides, $\gamma$ is \[\gamma := \sqrt{\frac{\det g}{\det G^{-1}}}.\]
For the sake of completeness, let us just glance at how $\mathcal{E}$ transforms a basis $(e_i \oplus \tilde{e}^i)$ into another basis:
\[
(e_i \oplus \tilde{e}^i) \mapsto e^{\kappa \phi} \big((e_i - (g-B)_{ij}\tilde{e}^j) \oplus \gamma((g+B)^{-1\; ij} e_j + \tilde{e}_i)\big)
\]
We will exclusively work with the basis induced on $TM \oplus T^* M$ by the local chart on $M$.
Notice moreover that if the bundle had an $O(d,d)$ structure group, then the vielbein would be reducing the structure group to $O(d)\times O(d)\times \mathbb{R}$.

Suppose that the vector bundle $TM \oplus T^* M$ encodes actually the standard exact Courant algebroid $(TM \oplus T^* M, \langle -, -\rangle , [-,-]_{\text{D}}, \rho)$ (see e.g.\,\cite{Roy}). If we take sections $U = X + \alpha$ and $V = Y + \beta$, the canonical choices for the pairing, the Dorfman bracket and the "anchor map" $\rho$ are:
\begin{align}
	\langle U, V \rangle & = \iota_X \beta+ \iota_Y \alpha, \\
	[U,V]_{\text{D}} &= [X,Y]_{\text{Lie}} + \mathcal{L}_{X}\beta - \iota_{Y}\mathrm{d} \alpha,  \\
	\rho(U)& =\text{pr}(U)=X .
\end{align}
Then $\mathcal{E}$ induces a Courant algebroid homomorphism with\\$(TM \oplus T^{*}M, [-,-]_{\text{D}}, \mathcal{G}, \rho)$, where $\mathcal{E}([U,V]_{\text{D}}) = [\mathcal{E} U, \mathcal{E}V]_{\text{D}}$,\footnote{More explicitly, the Dorfman bracket on the LHS is just written in the new basis.} and the pairing becomes 
\begin{equation}
	\mathcal{G} \equiv e^{2\kappa \phi} \begin{pmatrix} -2 g(x) & 0 \\ 0 & 2 G^{-1}(x) \gamma^2 \end{pmatrix},
\end{equation}
whereas \begin{equation} \rho(U) = e^{\kappa \phi}(X + \gamma (G^{-1}+\Pi)(\alpha)) \label{anchor} \end{equation} 
is the new anchor map.

Let us think of $\Gamma(TM \oplus T^{*}M)$ as a bimodule, for a Lie-like bracket $\llbracket-,-\rrbracket$ satisfying Jacobi identity, anti-symmetry and $\mathbb{R}$-linearity. Among all the possibilities, and in the basis induced by the coordinates on $M$, we choose it to be:
\begin{equation}
\llbracket U,V\rrbracket = \left(\rho(U)Y^i - \rho(V)X^i \right)\partial_i \oplus \left(\rho(U)\beta_i - \rho(V)\alpha_i \right) \mathrm{d}x^i.
\end{equation}
The map $\rho: \Gamma(TM \oplus T^{*}M) \mapsto \Gamma(TM)$ is the aforementioned anchor. Then an affine connection on the sections of the Courant algebroid, with completely skew torsion, and metric with respect to the Courant algebroid pairing, can be extracted in the following way, from the difference of the two brackets:

\begin{equation}
	\langle[U,V]_{\text{D}} - \llbracket U,V \rrbracket, W\rangle = \langle\nabla_W U, V\rangle, \quad W \in \Gamma(TM \oplus T^*M).
	\label{nab}
\end{equation}
Conventionally Courant algebroid connections are represented by $\nabla_{-}$ rather than $\nabla_{\rho(-)}$, i.e. it is usual to not display the anchor map.

The interesting non-flat connection arises when working in the $\mathcal{E}$ basis for $[-,-]_{\text{D}}$:
\[
\mathcal{G}\left([U,V]_{\text{D}} - \llbracket U, V \rrbracket, W\right) = \mathcal{G}\left(\nabla_{W}U,V \right).
\]
However we will not need a fully-fledged connection, as we want to look at the dual vector spaces $TM$ and $T^*M$, dual with respect to the canonical pairing of vector fields with $1$-forms. For a specific instance of $\kappa$ and dimension of the base manifold $M$, the $TM$ case was already studied in \cite{Boffo:2019zus}. We will comment on this later. Let us hence focus on $T^*M$. For our current purpose, we will inspect the short exact sequence:
\[
0 \mapsto TM \overset{\Delta}{\hookrightarrow} TM \oplus T^*M \overset{\Delta^*}{\twoheadrightarrow} T^*M \mapsto 0.
\]
Here, $\Delta$ embeds $TM$ into $TM \oplus T^{*}M$ in the following way:
\begin{equation}
	\Delta(X) =  \frac{1}{2} e^{-\kappa \phi} \left( g^{-1}(g+B)(Y) + \gamma^{-1} G(Y) \right).
\end{equation}
Using the Courant algebroid metric $\mathcal{G}$ to identify $TM \oplus T^*M$ with its dual, we get the surjective map $\Delta^*: \Gamma(TM \oplus T^*M) \twoheadrightarrow \Gamma(T^*M)$, which sends sections of the generalized tangent bundle into sections of the cotangent bundle in this way:
\[
\Delta^*(X+\alpha) = e^{\kappa\phi} \left( - (g-B)(X) + \gamma \alpha \right).
\]
A closer inspection should suffice to convince oneself that $\text{ker}(\Delta^*) = \text{im}(\Delta)$, as it should be. 

The short exact sequence can be split with the help of $r : \Gamma(T^*M) \mapsto \Gamma(TM \oplus T^*M)$,
\begin{equation}
	r(\alpha) = e^{-\kappa \phi} \gamma^{-1}\alpha.
	\label{r}
\end{equation}
Now one can demand to work with a connection on covectors by implementing only $r$-generalized vectors in the formula for the connection \eqref{nab}:
\begin{equation}
	\mathcal{G}\left([r(\alpha), r(\beta)] - \llbracket r(\alpha), r(\beta) \rrbracket, r(\nu)\right) \equiv \mathcal{G}(\nabla_r(\nu) r(\alpha), r(\beta)) =: 2 G^{-1}(\widetilde{\nabla}_{\rho(\nu)} \alpha, \beta).
\end{equation}
Here we defined $\widetilde{\nabla}: \Gamma(T^*M) \mapsto \Gamma( TM \otimes T^*M )$ as
\[
\nabla_{r(\nu)} r(\alpha) =: r(\widetilde{\nabla}_{\rho(\nu)} \alpha)\]
and used the induced metric $\mathcal{G}(r(-),r(-) ) = 2G^{-1}(-,-)$.

Now the splitting \eqref{r} combined with the Courant algebroid bracket will exactly forget all the scalar factors so that, upon a clever extraction of some terms resulting from $\llbracket r(\alpha), r(\beta)\rrbracket$, one obtains, combining all the expressions together, the connection for the case $\phi =0, \gamma =1$ in the coordinate basis:
\begin{align}
	2 G^{kl} \mathit{\Gamma}^{ij}{}_{k} = & \left(G^{-1}+\Pi\right)^{im}\partial_{m} \left(G^{-1}+\Pi\right)^{jl} \notag \\ \, & + 2\left(G^{-1}+\Pi\right)^{[j\vert m}\partial_m \left(G^{-1}+\Pi\right)^{\vert i] l}.
\end{align}
Of course this is not the end of the story, as some additional terms from the Lie-like bracket due to the derivative $\rho(r(-))$ hitting the scalar factors have not been unveiled yet. Let us denote
\begin{equation}
-\partial D_\kappa = e^{\kappa \phi} \gamma \partial (e^{-\kappa \phi} \gamma^{-1}) = -\kappa \partial \phi - \frac{1}{2} g^{ln} \partial g_{ln} + \frac{1}{2} G_{ln}\partial G^{ln}. 
\end{equation}
Eventually, in the holonomic coordinate basis, the connection coefficients can be checked to be:
\begin{equation}
	\mathit{\Gamma}^{ij}{}_{k} + \left(\delta^i_k \left(G^{-1}+\Pi\right)^{j m} -G_{pk}G^{ij}\left(G^{-1}+\Pi\right)^{p m} \right) \partial_{m}D_\kappa \equiv \mathit{\Gamma}^{ij}{}_{k} + T^{ij}{}_{k}. \label{con-1f}
\end{equation}
The partial derivatives are the result of differentiating $e^{\kappa \phi}$ and $\gamma$. For the sake of convenience the last two summands are called $\mathit{T}^{ij}{}_{k} \in C^{\infty}( \vee^2 TM \wedge T^*M)$. $\mathit{T}^{ij}{}_{k}$ is antisymmetric in $j,k$ and symmetric in $i,k$ and $j,i$ respectively.

\subsection{The curvature scalar}

A curvature scalar built from the Riemann curvature tensor $\text{Riem}$:
\[ \text{Riem}(U,V,W) = \left[\nabla_{U},\nabla_V\right] W - \nabla_{\llbracket U,V\rrbracket} W, \quad U,V,W \in \Gamma(TM \oplus T^*M)\]
yields an invariant Lagrangian. On the $r$-sections of course the Riemann curvature tensors for the connections $\nabla$ and $\widetilde{\nabla}$ are related by $r (\text{Riem}(U,V,W)) = \text{Riem}(r(U), r(V), r(W))$. We want to focus especially on the Ricci curvature tensor of the connection \eqref{con-1f} contracted with the non-symmetric combination $G^{-1}-\Pi$: 
\begin{equation}
	\text{Riem}_{m}{}^{lij} \delta^m_i G_{lp}\left(G^{-1}-\Pi\right)^{pq}G_{qj}.
	\label{Riem}
\end{equation} 
For the formula of the Riemann curvature due to $\mathit{\Gamma}^{ij}{}_{k}$ we refer to our previous paper \cite{Boffo:2020}. Now let us instead establish how $\mathit{T}^{ij}{}_{k}$ contributes. Its contribution can be written as:
\begin{align}
	-\mathit{T}^{mk}{}_{i} G_{mp} G_{qk} \widetilde{\nabla}^{i}_{\phi_0, \gamma_1} \left(G^{-1}-\Pi\right)^{pq} & \, \notag \\
	+ \big(\text{Tor}^{im}{}_{l} \mathit{T}^{lk}{}_{i} + 2 \mathit{T}^{[i\vert l}{}_{i} \mathit{T}^{\vert m]k}{}_{l}\big) & \left(G^{-1}-\Pi\right)^{pq}G_{mp} G_{qk}
	\label{new-part}
\end{align}
In the above expression, $\widetilde{\nabla}^{i}_{\phi_0, \gamma_1}$ is the connection when $\phi=0, \gamma =1$, which has the totally antisymmetric torsion $\text{Tor}^{ijk}$:

\begin{equation}
\text{Tor}^{ijk} =	2 \mathcal{R}^{ijk} + 2 \mathcal{Q}^{ij}{}_{l} G^{lk} + 4 G^{r [i\vert}\mathcal{Q}^{\vert j] k}{}_{r}.
\end{equation}
Here we have introduced the fluxes in the Supergravity frame (which is the frame with metric $g$, $2$-form $B$ and $\phi \in C^\infty (M)$ solving the bosonic part of the Supergravity action). Their local expressions are:
\begin{align}
	\mathcal{R}^{ijk} := & 3 \Pi^{[i\vert m}\partial_m \Pi^{\vert jk]},\\
	\mathcal{Q}^{ij}{}_{k} := & \partial_k \Pi^{ij}.
\end{align}

For the first term in \eqref{new-part} we have integrated by part against the volume form and used that $\widetilde{\nabla}^{i}_{\phi_0, \gamma_1}$ is a metric connection for $G$. Then $T^{mki} = - T^{mik}$ dictates that the covariant derivative of $\Pi$ must be completely antisymmetrized\footnote{With a very little effort one can notice that 
	\[ \widetilde{\nabla}^{[i}_{\phi_0, \gamma_1} \Pi^{pq]} = \partial^{[i} \Pi^{pq]} + \text{Tor}^{ip}{}_{l} \Pi^{lq} -\text{Tor}^{iq}{}_{l} \Pi^{lp} + \text{Tor}^{pq}{}_{l} \Pi^{li}. \]}. However, since $T$ is symmetric in the first two slots, this term drops out of the expression. 

The second term in \eqref{new-part} is contracted just with the bivector. Disclosing the torsion in its components, the final result is:
\begin{equation}
	 2\left(G^{-1}+\Pi\right)^{st} \partial_{t} D_\kappa \left[G_{si} ( \mathcal{R}^{imk} - \mathcal{Q}^{imk}+ \mathcal{Q}^{ikm})\Pi_{mk} + \mathcal{Q}^{mk}{}_{s} \Pi_{mk}\right].
	\label{e1}
\end{equation}

As of the remaining bit, it yields the following:
\begin{equation}
	(1-d)(d-2) \left((G^{-1}+\Pi)^{ki}\partial_i D_\kappa \right)^{2}
	\label{e2}
\end{equation}

We can now combine all the pieces together and couple the curvature scalar to $ e^{-2\tilde{\phi}}\sqrt{\det G^{-1}}$. In doing so, one must not forget about integration by parts with $\widetilde{\nabla}_{\phi_0,\gamma_1}$, as this hits $e^{-2\tilde{\phi}}$. Before going on with the result, let us follow the strategy of our work \cite{Boffo:2019zus}: there, the kinetic term for the dilaton in 10 dimensions turned out to be accounted by a conformal factor of $e^{-2\phi/3}$, which rescaled the Riemannian metric $g$. Thus replacing $G \mapsto e^{2\tilde{\phi}/3}G$ twice in \eqref{Riem} and taking care of the square root of its  determinant, but still displaying $d$ dimensions and $\kappa$ factors though these are actually fixed:
\[ e^{-2\tilde{\phi}} \sqrt{\det G^{-1}} \equiv e^{(-4 \kappa + d\kappa)\tilde{ \phi}} \sqrt{\det G^{-1}} .\]
Furthermore, $ \partial D \equiv e^{\kappa \tilde{\phi}} \partial e^{-\kappa \tilde{\phi}} = - \kappa \partial \tilde{\phi}$, so we get:
	\begin{equation}
		-\kappa^2 (1-d)(d-4) \left( (G^{-1}+\Pi)^{ki}\partial_i \tilde{\phi} \right)^2 .
		\label{e3}
	\end{equation}

Putting \eqref{e1}, \eqref{e2} and \eqref{e3} together in the action $S$, we get
\begin{align}
	S = \int \mathrm{d}x \, &\sqrt{\det G^{-1}} e^{(d-4) \kappa \tilde{\phi}} \big[ \kappa^2 (1-d)(d-2-d+4)  \left( (G^{-1}+\Pi)^{ki}\partial_i \tilde{\phi} \right)^2 \notag \\
	& - 2 \kappa \left(G^{-1}+\Pi\right)^{st} \partial_{t} \tilde{\phi} \left[G_{si} ( \mathcal{R}^{imk} - 2\mathcal{Q}^{imk})\Pi_{mk} + \mathcal{Q}^{mk}{}_{s} \Pi_{mk}\right] \notag \\
	& + \text{R}_{G} - \frac{1}{12} \mathcal{R}^{2} - \dfrac{1}{2} \mathcal{R}^{lmi} \mathcal{Q}^{jk}{}_{i} G_{lj}G_{mk} - \dfrac{1}{4} \mathcal{Q}^{jl}{}_{m} \mathcal{Q}^{k n}{}_{i} G_{jk} G_{ln} G^{mi} \notag \\
	\, & -\dfrac{1}{2} \mathcal{Q}^{lj}{}_{k} \mathcal{Q}^{k m}{}_{l} G_{jm} \big].
	\label{act} 
\end{align}
In this formula, $\text{R}_{G}$ is the Ricci curvature scalar of the symmetric part of the connection. Remarkably, compared to literature on the topic, it does not simply come from a tensor constructed with the $G$ metric only, but it entails the bivector $\Pi$, because of the anchor map \eqref{anchor}.

In particular, in $10$ dimensions and when $\kappa = -1/3$, the kinetic term for the dual dilaton is:
\[
-2 \left((G^{-1}+\Pi)^{ki}\partial_i \tilde{\phi}\right)^2.
\]

\section{Conclusion}
In this short note we suggested a way to embed the dual dilaton in the construction of an invariant action for the open strings metric and the T-dual fluxes $\mathcal{R}$ and $\mathcal{Q}$. Our findings complement our previous works on the same topic \cite{Boffo:2019zus} \cite{Boffo:2020}. More importantly, one of the features of our particular ansatz is that it encompasses the effective closed string action for the NS-NS fields: This is obtained when one deploys a splitting $s: \Gamma(TM) \mapsto \Gamma(TM \oplus T^*M)$, $\rho \circ s = \text{id}_{TM}$ to focus on vector fields in \eqref{nab}, and then builds the curvature invariant according to the same technique.

Thus with the present work we think to have provided a new clear algebraic and geometric picture for the background fields for strings and the local fluxes.

\begin{acknowledgement}
Thanks to prof.\,Peter Schupp from Jacobs University Bremen for bringing the dual dilaton in gravity with fluxes to my attention. Thanks to dr.\,Afentoulidis Almpanis for discussion. A warm thanks to the organizer of "Lie Theory and its Applications in Physics" 2021 (LT-14) for this very interesting workshop oriented at strengthening the bonds between representation theory, algebra and geometry with mathematical physics. I acknowledge GA\v{C}R EXPRO 19-28628X for financial support.
\end{acknowledgement}

\begin{thebibliography}{99.}%
%
%


%

\bibitem{Andriot:2013xca} D. Andriot, A. Betz, \textit{JHEP},  \textbf{12} (2013) 083.

\bibitem{Andriot:2012an} D. Andriot, O. Hohm, M. Larfors,
D. Lust, P. Patalong, \textit{Fortsch. Phys.}, \textbf{60} (2012) 1150-1186.
	
\bibitem{Andriot:2012wx} D. Andriot, O. Hohm, M. Larfors,
	D. Lust, P. Patalong, \textit{Phys. Rev. Lett.} \textbf{108} (2012). 

\bibitem{Blumenhagen:2012nk} R. Blumenhagen, A. Deser, E. Plauschinn, F. Rennecke, \textit{Phys. Lett. B}, \textbf{720} (2013) 215-218.
	
\bibitem{Blumenhagen:2013aia} R. Blumenhagen, A. Deser, E. Plauschinn, F. Rennecke, C. Schmid, \textit{Fortsch. Phys.}, \textbf{61} (2013) 893-925.
	
\bibitem{Boffo:2019zus} E. Boffo, P. Schupp, \textit{JHEP}, \textbf{01} (2020) 007.

\bibitem{Boffo:2020} E. Boffo, P. Schupp, to be published in \textit{JHEP}, arXiv:2106.09601. 
	
\bibitem{Coimbra:2011nw} A. Coimbra, C. Strickland-Constable, D.
		Waldram, \textit{JHEP},  \textbf{11} (2011) 091.
		
\bibitem{Gualtieri2004} M. Gualtieri, 2011, \textit{Annals of Mathematics}, \textbf{174(1)} (2011) 75-123. \url{http://www.jstor.org/stable/23030560}

\bibitem{Moser1965} J. Moser, doi: 10.1090/s0002-9947-1965-0182927-5.
	
\bibitem{Roy} D. Roytenberg, A. Weinstein, \textit{Lett. Math. Physics}, \textbf{46} (1998) 81-93.	
%
%
\end{thebibliography}
\end{document}